# An example of a violation of the spatial quantum inequality with a comment on the quantum interest conjecture.

Dan Solomon
Rauland-Borg Corporation
Mount Prospect, IL
Email: dan.solomon@rauland.com
June 21, 2011

**Abstract.**

It is generally known that the energy density can be negative in quantum field theory. It is also believed that there are limits on this negative energy density. These limits are known as the quantum inequalities. In a recent paper [8] an example was provided of a system which violated the quantum inequalities. Here we will demonstrate a violation of the spatial quantum inequality for a scalar field with zero mass in 1-1 dimensional space-time. In addition it will be argued that the system presented here also violates the quantum interest conjecture.

## 1. Introduction.

It has been shown [1] that in quantum field theory the energy density for a free field can be negative over some region of space and time. This is a quantum effect that differs from classical physics where the energy density of a field is always positive. It has been suggested the existence of negative energy density could lead to certain "exotic" effects such as the existence of wormholes in space-time [2] and violations of the second law of thermodynamics [3]. However, a number of papers have been written which demonstrate the existence of the quantum inequalities (see [4-7] and references, therein). The quantum inequalities are lower limits on the weighted average of the energy density and their existence reduces the possibility that the aforementioned "exotic" effects occur.

In contradiction to the claims of the quantum inequalities the author has written several papers that claim to demonstrate violations of the quantum inequalities [8-11]. In this discussion we will demonstrate a counter example to the spatial quantum inequality.

For a scalar field with zero mass in 1-1 dimensional space-time in a free field (i.e. all external potentials are zero) the spatial quantum inequality is given by E. E. Flanagan [5] as,

$$\int_{-\infty}^{+\infty} T_{00R}(x,t)\rho(x)dx \geq \xi_{S,\min}[\rho] \tag{1.1}$$

where $T_{00R}(x,t)$ is the regularized energy density, $\rho(x)$ is a non-negative function, and $\xi_{S,\min}[\rho]$ is given by,

$$\xi_{S,\min}[\rho] = -\frac{1}{24\pi}\int_{-\infty}^{+\infty} dx \frac{\rho'(x)^2}{\rho(x)} \tag{1.2}$$

where $\rho'(x) \equiv d\rho(x)/dx$.

Consider a zero mass scalar field in the presence of a static scalar potential $V(x)$ where $V(x)$ is the inverse square well potential defined by,

$$V(x) = \begin{cases} V_0 & \text{for } |x| < a \\ 0 & \text{for } |x| > a \end{cases} \tag{1.3}$$

where $V_0 > 0$. In Section 2 the renormalized kinetic energy density $T_{00R}(x)$ will be calculated for this situation. It will be shown that for $|x| > a$, outside the inverse potential well, $T_{00R}(x)$ is zero and for $|x| < a$ the average value of the kinetic energy density is negative. It would be easy to show that this result violates the spatial quantum inequality, except for the fact that the spatial quantum inequality only applies to free fields, that is, quantum fields where the scalar potential is zero. In order to make the spatial quantum inequality apply to this case the scalar potential must be set to zero. The way to achieve this is to instantaneously remove the scalar potential. Assume that this occurs at time $t = 0$ so that the scalar potential is given by,

$$V(x,t) = \theta(-t)V(x) \tag{1.4}$$

where $\theta(u)$ is the Heaviside step function. What, then, happens to the kinetic energy density at $t=0$? As has been demonstrated in [8] and [10] the kinetic energy density is continuous at $t=0$. This fact shouldn't be surprising because it is consistent with classical physics. In classical physics an abrupt change in the external potential does not produce an abrupt change in the kinetic energy. So the quantum mechanical result is simply a carry over of the same result from classical physics [8].

Therefore, we know the kinetic energy density immediately after the scalar potential has been removed at $t = 0^+$. The final result is that at $t = 0^+$ we have a free quantum field for which the average kinetic energy density is negative in the region $-a < x < +a$ and zero outside of this region. Note that due to the fact that this is a free field (the scalar potential has been set to zero) the kinetic energy density is the same as the energy density. It is then easy to show that the spatial quantum inequality will be violated. This will be done in Section 4. Also, in Section 6, we will examine the case were the scalar potential is reduced to zero in a continuous manner.

## 2. Kinetic energy density.

In this section we will calculate the kinetic energy density for a scalar field with zero mass in the presence of scalar potential given by Eq. (1.3). In making this calculation we will use some of the results from a paper by N. Graham and K.D Olum [12]. They calculated the energy density for a scalar field in the presence of an inverse square well scalar potential in 2-1 dimensional space-time. Their approach will be used and appropriately modified to the problem here which is to calculate the kinetic energy density of scalar field in 1-1 dimensional space-time.

The field operator $\hat{\varphi}_\lambda(x,t)$ satisfies the Klein-Gordon equation,

$$\frac{\partial^2 \hat{\varphi}_\lambda}{\partial t^2} - \frac{\partial^2 \hat{\varphi}_\lambda}{\partial x^2} + \lambda V(x)\hat{\varphi}_\lambda = 0 \tag{2.1}$$

where $\lambda$ is a positive parameter that can be set to zero to turn off the potential and $V(x)$ is given by (1.3). The kinetic energy density operator is,

$$\hat{T}_{00}[\hat{\varphi}_\lambda] = \frac{1}{2}\left(\frac{\partial \hat{\varphi}_\lambda}{\partial t} \cdot \frac{\partial \hat{\varphi}_\lambda}{\partial t} + \frac{\partial \hat{\varphi}_\lambda}{\partial x} \cdot \frac{\partial \hat{\varphi}_\lambda}{\partial x}\right) \tag{2.2}$$

Note the energy density operator, $\xi[\hat{\varphi}_\lambda]$, is related to the kinetic energy density operator by,

$$\xi[\hat{\varphi}_\lambda] = \hat{T}_{00}[\hat{\varphi}_\lambda] + \frac{1}{2}\lambda V(x)\hat{\varphi}_\lambda \cdot \hat{\varphi}_\lambda \tag{2.3}$$

The difference between the two operators is that the energy density operator has an explicit dependence on the scalar potential. If the scalar potential is zero the kinetic energy density and energy density are the same.

The field operator is given by,

$$\hat{\varphi}_\lambda(x,t) = \int_0^\infty \frac{dk}{\sqrt{2\pi\omega}} \sum_{\chi=+,-} \left( \hat{a}_{\lambda,k}^\chi f_{\lambda,k}^\chi(x,t) + \hat{a}_{\lambda,k}^{\chi*} f_{\lambda,k}^{\chi*}(x,t) \right) \tag{2.4}$$

where,

$$f_{\lambda,k}^\chi(x,t) = \psi_{\lambda,k}^\chi(x) e^{-i\omega t} \tag{2.5}$$

and where $\omega = |k|$ and the $\psi_{\lambda,k}^\chi(x)$ are real with $\psi_{\lambda,k}^+(x)$ representing the symmetric solutions and $\psi_{\lambda,k}^-(x)$ representing the anti-symmetric solutions. The $\hat{a}_{\lambda,k}^{\chi*}$ and $\hat{a}_{\lambda,k}^\chi$ are the creation and destruction operators, respectively. They obey the commutation relationships $\left[ \hat{a}_{\lambda,k'}^{\chi'}, \hat{a}_{\lambda,k}^{\chi*} \right] = \delta_{\chi'\chi} \delta(k'-k)$ with all other commutations being zero. The $\psi_{\lambda,k}^\chi(x)$ are solutions of,

$$-k^2 \psi_{\lambda,k}^\chi(x) - \frac{d^2 \psi_{\lambda,k}^\chi(x)}{dx^2} + \lambda V(x) \psi_{\lambda,k}^\chi(x) = 0 \tag{2.6}$$

where,

$$\int_{-\infty}^{+\infty} dx\, \psi_{\lambda,k}^+(x) \psi_{\lambda,k'}^-(x) = 0 \text{ and } \sum_{\chi=+,-} \int_{-\infty}^{+\infty} \psi_{\lambda,k}^\chi(x) \psi_{\lambda,k'}^\chi(x) dx = 2\pi \delta(k-k') \tag{2.7}$$

The above relationships define the field operator. However, in order to completely specify the system, the state vector on which the field operator acts must also be defined. The normalized state vector will be designated by $|0_\lambda\rangle$ and obeys the relationship $\hat{a}_{\lambda,k} |0_\lambda\rangle = 0$.

From the above discussion the kinetic energy density expectation value is,

$$T_{00,\lambda} = \langle 0_\lambda | \hat{T}_{00}[\hat{\varphi}_\lambda] | 0_\lambda \rangle \tag{2.8}$$

Using the above relationships we obtain,

$$T_{00,\lambda}(x) = \int_0^\infty \left( \frac{dk}{2\pi |k|} \right) \sum_{\chi=+,-} \frac{1}{2} \left[ \left| \frac{\partial f_{\lambda,k}^\chi(x,t)}{\partial t} \right|^2 + \left| \frac{\partial f_{\lambda,k}^\chi(x,t)}{\partial x} \right|^2 \right] dk \tag{2.9}$$

Next, use (2.5) in the above to obtain,

$$T_{00,\lambda}(x) = \int_0^\infty \left(\frac{dk}{2\pi|k|}\right) \sum_{\chi=+,-} \frac{1}{2}\left[k^2 \psi_{\lambda,k}^{\chi}(x)^2 + \left(\frac{d\psi_{\lambda,k}^{\chi}(x)}{dx}\right)^2\right] dk \qquad (2.10)$$

In order to regularize this solution we proceed as in Ref. [12] and introduce a counter term to compensate for the cosmological constant. This involves subtracting off the term,

$$T_{00,0}(x) = \int_0^\infty \left(\frac{dk}{2\pi|k|}\right) \sum_{\chi=+,-} \frac{1}{2}\left[k^2 \psi_{0,k}^{\chi}(x)^2 + \left(\frac{d\psi_{0,k}^{\chi}(x)}{dx}\right)^2\right] dk \qquad (2.11)$$

where,

$$\psi_{0,k}^{+}(x) = \cos(kx) \text{ and } \psi_{0,k}^{-}(x) = \sin(kx) \qquad (2.12)$$

Therefore the regularized kinetic energy density is,

$$T_{00R}(x) = \int_0^\infty \left(\frac{dk}{4\pi|k|}\right) \sum_{\chi=+,-} \left[k^2\left(\psi_{\lambda,k}^{\chi}(x)^2 - \psi_{0,k}^{\chi}(x)^2\right) + \left(\frac{d\psi_{\lambda,k}^{\chi}(x)}{dx}\right)^2 - \left(\frac{d\psi_{0,k}^{\chi}(x)}{dx}\right)^2\right] \qquad (2.13)$$

This can be rewritten using the relationship,

$$\left(\frac{d\psi(x)}{dx}\right)^2 = \frac{1}{2}\frac{d^2}{dx^2}\psi(x)^2 - \psi(x)\frac{d^2\psi(x)}{dx^2} \qquad (2.14)$$

Use (2.6) in the above to obtain,

$$\left(\frac{d\psi_{\lambda,k}^{\chi}(x)}{dx}\right)^2 = \frac{1}{2}\frac{d^2}{dx^2}\psi_{\lambda,k}^{\chi}(x)^2 + \left(k^2 - \lambda V(x)\right)\psi_{\lambda,k}^{\chi}(x)^2 \qquad (2.15)$$

Use this in (2.13) to obtain,

$$T_{00R}(x) = \int_0^\infty \left(\frac{dk}{4\pi|k|}\right) \sum_{\chi=+,-} \begin{bmatrix} 2k^2\left(\psi_{\lambda,k}^{\chi}(x)^2 - \psi_{0,k}^{\chi}(x)^2\right) \\ +\left(\frac{1}{2}\frac{d^2}{dx^2}\psi_{\lambda,k}^{\chi}(x)^2 - \lambda V(x)\psi_{\lambda,k}^{\chi}(x)^2\right) \\ -\left(\frac{1}{2}\frac{d^2}{dx^2}\psi_{0,k}^{\chi}(x)^2\right) \end{bmatrix} \qquad (2.16)$$

From [12] we can relate the mode solutions to the Green's function,

$$\sum_{\chi=+,-} \psi_{\lambda,k}^{\chi}(x)^2 = 2k\,\text{Im}\,G_\lambda(x,x,k) \qquad (2.17)$$

where $G_\lambda(x,x',k)$ is the Green's function, which satisfies,

$$-G_\lambda''(x,x',k) + \lambda V(x) G_\lambda(x,x',k) - k^2 G_\lambda(x,x',k) = \delta(x-x') \tag{2.18}$$

From [12],

$$G_0(x,x,k) = \frac{i}{2k} \tag{2.19}$$

Use this and (2.17) in (2.16) to obtain,

$$T_{00R}(x) = \int_0^\infty \left(\frac{2k\,dk}{4\pi|k|}\right) \text{Im}\left[\left(2k^2 G_\lambda(x,x,k) - ik\right) + \left(\frac{1}{2}\frac{d^2}{dx^2} G_\lambda(x,x,k) - \lambda V(x) G_\lambda(x,x,k)\right)\right] \tag{2.20}$$

From [12] we have the relationship $G_\lambda(x,x',k) = G_\lambda(x,x',-k^*)^*$. Therefore for real $k$ we have,

$$\text{Im}\, G_\lambda(x,x,k) = \frac{1}{2i}\left(G_\lambda(x,x,k) - G_\lambda(x,x,-k)\right) \tag{2.21}$$

Use this in (2.20) to obtain,

$$T_{00R}(x) = \int_{-\infty}^\infty \left(\frac{2k\,dk}{8\pi i|k|}\right)\left[\left(2k^2 G_\lambda(x,x,k) - ik\right) + \left(\frac{1}{2}\frac{d^2}{dx^2} G_\lambda(x,x,k) - \lambda V(x) G_\lambda(x,x,k)\right)\right] \tag{2.22}$$

Note that there are no poles in the upper half of the complex plane [12]. Therefore we can deform the integration path in the upper half plane per [12] and integrate around the branch cut along the positive imaginary axis due to $|k|$ to obtain,

$$T_{00R}(x) = -\frac{1}{4\pi}\int_0^\infty d\kappa \left[\left(4\kappa^2 G_\lambda(x,x,i\kappa) - 2\kappa\right) - \left(\frac{d^2}{dx^2} G_\lambda(x,x,i\kappa) - 2\lambda V(x) G_\lambda(x,x,i\kappa)\right)\right] \tag{2.23}$$

Outside the square well potential where $|x| > a$ it is shown in [12] that,

$$G_\lambda(x,x,i\kappa) = \frac{1 + r(i\kappa)e^{-2\kappa x}}{2\kappa} \tag{2.24}$$

where,

$$r(i\kappa) = -\frac{\lambda V_0 e^{2\kappa a} \tanh(2\kappa' a)}{2\kappa'\kappa + (\kappa^2 + \kappa'^2)\tanh(2\kappa' a)} \tag{2.25}$$

and where $\kappa' = \sqrt{\kappa^2 + \lambda V_0}$. Use (2.24) and (2.25) in (2.23) along with $V(x) = 0$ for $|x| > a$ to obtain,

$$T_{00R}(|x| > a) = 0 \tag{2.26}$$

According to [12] the Green's function for $|x| < a$ is,

$$G_\lambda(x, x, i\kappa)\bigg|_{|x|<a} = \frac{N_\kappa + \lambda V_0 \cosh(2\kappa' x)}{2\kappa' D_\kappa} \tag{2.27}$$

where,

$$N_\kappa = (\kappa^2 + \kappa'^2)\cosh(2\kappa' a) + 2\kappa' \kappa \sinh(2\kappa' a) \tag{2.28}$$

and,

$$D_\kappa = 2\kappa' \kappa \cosh(2\kappa' a) + (\kappa^2 + \kappa'^2)\sinh(2\kappa' a) \tag{2.29}$$

Next calculate the total kinetic energy in the region $-a < x < a$. This is given by,

$$E_{KE} = \int_{-a}^{+a} T_{00R}(x) dx \tag{2.30}$$

Use (2.23) in the above to obtain,

$$E_{KE} = -\frac{1}{4\pi}\int_0^\infty d\kappa \left[ \begin{array}{c} (4\kappa^2 F_\lambda(\kappa) - 4\kappa a) + 2\lambda V_0 F_\lambda(\kappa) \\ -\left(\frac{dG_\lambda(x,x,i\kappa)}{dx}\bigg|_{x=a} - \frac{dG_\lambda(x,x,i\kappa)}{dx}\bigg|_{x=-a}\right) \end{array} \right] \tag{2.31}$$

where,

$$F_\lambda(\kappa) = \int_{-a}^{+a} G_\lambda(x, x, i\kappa) dx = \frac{2aN_k + \lambda V_0 \sinh(2a\kappa')/\kappa'}{2\kappa' D_\kappa} \tag{2.32}$$

and,

$$\frac{d}{dx}G_\lambda(x, x, i\kappa)\bigg|_{|x|<a} = \frac{\lambda V_0 \sinh(2\kappa' x)}{D_\kappa} \tag{2.33}$$

Using (2.33) and (2.32) in (2.31) it is shown in the Appendix that,

$$E_{KE} = -\frac{V_0^2}{4\pi}\int_0^\infty d\kappa \left\{ \frac{2a\kappa' \cosh(2a\kappa') - \sinh(2a\kappa')}{\kappa'^2 D_\kappa} \right\} \tag{2.34}$$

where we have set $\lambda = 1$. The integrand is non-negative for all $\kappa \geq 0$. Therefore $E_{KE} < 0$. The result of this is that the total kinetic energy is negative in the interval $-a < x < a$ and the kinetic energy density is zero outside of this region.

## 3. Removing the potential.

At this point we have calculated the kinetic energy density for an inverse square well potential. In order to use these results to test the validity of the quantum inequalities the potential must be removed due to the fact that the quantum inequalities only apply to systems where external potential is zero. In Section 6 we will examine the situation where the scalar potential is removed continuously. In the present section we will consider the simpler case of an instantaneous removal of the potential. Therefore we assume that the potential takes the form of Eq. (1.4), i.e. $V(x,t) = \theta(-t)V(x)$ where $V(x)$ is given by (1.3). From the previous section we know what the kinetic energy density is during the period of time $t < 0$. What is the effect on the kinetic energy density due to the instantaneous removal of the scalar potential at time $t = 0$?

As has been discussed elsewhere [8][10] the kinetic energy density is continuous with respect to an abrupt change of the scalar potential. This result should not be surprising or unexpected because it is consistent with classical physics. This has been shown in Ref. [8] where the effect on the kinetic energy of a classical system due an instantaneous change in the potential was examined. It was shown that the kinetic energy was continuous in this case. This was also shown to be true for the quantum system under consideration here which will be demonstrated as follows.

Assume that the scalar potential is given by $V(x,t) = \theta(-t)V(x)$. In this case the equations of motion for the field operator are,

$$\frac{\partial^2 \hat{\varphi}}{\partial t^2} - \frac{\partial^2 \hat{\varphi}}{\partial x^2} + V(x)\hat{\varphi} = 0 \text{ for } t < 0 \qquad \frac{\partial^2 \hat{\varphi}^{(+)}}{\partial t^2} - \frac{\partial^2 \hat{\varphi}^{(+)}}{\partial x^2} = 0 \text{ for } t > 0 \qquad (3.1)$$

where $\hat{\varphi}^{(+)}(x,t)$ is the field operator for $t > 0$. The boundary condition at $t = 0$ are,

$$\hat{\varphi}^{(+)}(x,0) = \hat{\varphi}(x,0) \text{ and } \frac{\partial \hat{\varphi}^{(+)}(x,0)}{\partial t} = \frac{\partial \hat{\varphi}(x,0)}{\partial t} \qquad (3.2)$$

Use this in (2.2) to obtain,

$$\hat{T}_{00}^{(+)}(x,0) = \hat{T}_{00}(x,0) \tag{3.3}$$

where $\hat{T}_{00}^{(+)}(x,0) = \hat{T}_{00}\left[\hat{\varphi}^{(+)}(x,0)\right]$. Therefore the kinetic energy density operator is continuous at $t = 0$.

We can also expand $\hat{\varphi}^{(+)}(x,t)$ in terms of mode solutions to obtain,

$$\hat{\varphi}_\lambda^{(+)}(x,t) = \int_0^\infty \frac{dk}{\sqrt{2\pi\omega}} \sum_{\chi=+,-} \left(\hat{a}_{\lambda,k}^\chi f_{\lambda,k}^{\chi(+)}(x,t) + \hat{a}_{\lambda,k}^{\chi*} f_{\lambda,k}^{\chi(+)*}(x,t)\right) \tag{3.4}$$

Using (3.2) the initial condition, at $t = 0$, for the $f_{\lambda,k}^{\chi(+)}(x,t)$ are,

$$f_{\lambda,k}^{\chi(+)}(x,0) = f_{\lambda,k}^\chi(x,0) \text{ and } \frac{\partial f_{\lambda,k}^{\chi(+)}(x,0)}{\partial t} = \frac{\partial f_{\lambda,k}^\chi(x,0)}{\partial t} \tag{3.5}$$

This can be used in (2.9) to show that the kinetic energy density is continuous at $t = 0$. All this is discussed in more detail in [8].

At $t = 0$ we have removed the scalar potential so that the spatial quantum inequality should apply for $t = 0^+$. Since the kinetic energy density is continuous at $t = 0$ we know the value of the kinetic energy density just after the potential is removed. We have $T_{00R}^{(+)}(x,0) = T_{00R}(x)$ where $T_{00R}(x)$ is the kinetic energy density that was calculated in Section 2 for the inverse square well potential. In addition, since the potential is zero, the kinetic energy density is now equivalent to the energy density. Therefore the spatial quantum inequality should apply to $T_{00R}(x)$.

## 4. Violating the quantum inequality.

In this section we will show that the quantity $T_{00R}(x)$ violates the spatial quantum inequality. Recall from Section 2 that we have defined the quantity $E_{KE}$ as the total kinetic energy within the region $x < |a|$ (see Eq. (2.30)). It has been shown that $E_{KE} < 0$. In addition the kinetic energy density $T_{00R}(x)$ outside of the region $x > |a|$ is zero. To show that this violates the spatial quantum inequality let the weighting function $\rho(x)$ be given by,

$$\rho(x) = N \left\langle \begin{array}{l} 1 \text{ for } |x| < a \\ \exp\left[\eta(a-|x|)\right] \text{ for } |x| > a \end{array} \right. \tag{4.1}$$

where $N = \eta/[2(a\eta+1)]$. Use this in (1.2) to obtain,

$$\xi_{S,\min}[\rho] = -\frac{1}{12\pi}\eta N \tag{4.2}$$

Use these results along with those of the last section to obtain,

$$\int_{-\infty}^{+\infty} T_{00R}^{(+)}(x,0)\rho(x)dx = \int_{-a}^{+a} T_{00R}(x,0)dx = E_{KE}N = -|E_{KE}|N \tag{4.3}$$

The last step is allowed because $E_{KE}$ is negative (see Eq. (2.34). Also, in evaluating this integral we have used the fact that $T_{00,R}(x) = 0$ for $|x| > a$. Next, use (4.2) and (4.3) in Eq. (1.1) to yield,

$$-|E_{KE}|N \geq -\frac{1}{12\pi}\eta N \tag{4.4}$$

If the spatial quantum inequality is true this equation must be valid for all $\eta$. It is easy to show that this is not the case. Rewrite this result as,

$$\frac{1}{12\pi}\eta \geq |E_{KE}| \tag{4.5}$$

In the limit $\eta \to 0$ the left hand side of this equation approaches zero and the right hand side is unchanged. In this limit (4.5) is not true so that the spatial quantum inequality is not valid.

## 5. The positively of the energy.

There is one potential problem with the above results which will be addressed in this section. Consider the situation for $t = 0^+$ just after the scalar potential has been removed. The energy density at a given point is either negative or zero. Therefore the total energy integrated over all space is negative. This cannot be correct because the total energy cannot be less than or equal to zero.

The solution to the problem is given in a discussion in Ref. [8] and will be briefly reviewed here. Assume at some initial time $t_1$ a system is in its initial unperturbed vacuum state with $V(x) = 0$. The mode solutions for the field operator defined in Eq. (2.4) are given by $f_{0,k}^{\chi}(x,t) = \psi_{0,k}^{\chi}(x)e^{-i\omega t}$ where the $\psi_{0,k}^{\chi}(x)$ are the unperturbed solutions given by (2.12). Between time $t_1$ and $t_2$ the scalar potential is applied. At $t_2$ is reaches is final value which as specified in (1.3).

In the process of establishing the scalar potential a disturbance is created which moves outward at the speed of light. This perturbs the mode solutions $f_k^\chi(x,t)$. After a sufficiently long time $t_f \gg t_2$ the mode solutions will be approach their final value $f_{\lambda,k}^\chi(x,t_f) = \psi_{\lambda,k}^\chi(x) e^{-i\omega t_f}$ over some large region $-L \leq x \leq +L$.

This resolves the problem posed at the beginning at this section. Consider the region $-L \leq x \leq +L$ where $L \to \infty$ for sufficiently large $t_f \gg t_2$. Inside of this region the total kinetic energy is negative and is equal to the result calculated in Section 2. However outside of this region the kinetic energy density will be positive which will make the total integrated kinetic energy over all space positive as required. We are justified in ignoring this part of the solution because it is at infinity and therefore doesn't affect the result of our integrations over the sampling functions.

## 6. Time dependent change in potential.

In Section 3 it was argued that the kinetic energy density was continuous despite an instantaneous change in the scalar potential. In this section we will consider what happens if the scalar potential is reduced to zero in a continuous fashion.

We will start by setting up the problem in the Schrödinger picture. In this case the Schrödinger picture field operators $\hat{\varphi}_S(x)$ and $\hat{\pi}_S(x)$ are constant in time and the time dependence is associated with the Schrödinger picture state vector $|\Omega_S(t)\rangle$. In the Schrödinger picture the kinetic energy density operator is given by,

$$\hat{T}_{00S}(x) = \frac{1}{2}\left(\pi_S(x)^2 + \left(\frac{d\hat{\varphi}_S(x)}{dx}\right)^2\right) \qquad (6.1)$$

The state vector obeys the Schrödinger equation,

$$i\frac{\partial}{\partial t}|\Omega_S(t)\rangle = \hat{H}_S |\Omega_S(t)\rangle; \quad -i\frac{\partial}{\partial t}\langle\Omega_S(t)| = \langle\Omega_S(t)|\hat{H}_S \qquad (6.2)$$

$\hat{H}_S$ is the Schrödinger picture Hamiltonian operator and is given by,

$$\hat{H}_S = \hat{H}_{SKE} + \hat{H}_{SV} \qquad (6.3)$$

where,

$$\hat{H}_{SKE} = \int \hat{T}_{00S}(x)dx \qquad (6.4)$$

and,

$$\hat{H}_{SV} = \frac{1}{2}\int V(x,t)\hat{\varphi}_S(x)^2 dx \qquad (6.5)$$

Note that the $\hat{H}_{SKE}$ term in the Hamiltonian may be thought of as an operator corresponding to the total kinetic energy and $\hat{H}_{SV}$ is that part of the Hamiltonian that explicitly depends on the potential. With this in mind define the total kinetic energy as,

$$E_K(t) = \langle \Omega_S(t)|\hat{H}_{SKE}|\Omega_S(t)\rangle - \xi_K \qquad (6.6)$$

where $\xi_K$ is a renormalization constant. Take the derivative of (6.6) with respect to time to obtain,

$$\frac{d}{dt}E_K(t) = i\langle \Omega_S(t)|[\hat{H}_S, \hat{H}_{SKE}]|\Omega_S(t)\rangle \qquad (6.7)$$

Use (6.3) in the above to obtain,

$$\frac{d}{dt}E_K(t) = i\langle \Omega_S(t)|[\hat{H}_S, \hat{H}_S - \hat{H}_{SV}]|\Omega_S(t)\rangle = -i\langle \Omega_S(t)|[\hat{H}_S, \hat{H}_{SV}]|\Omega_S(t)\rangle \qquad (6.8)$$

Use (6.5) and (6.2) in the above to obtain,

$$\frac{d}{dt}E_K(t) = -\frac{1}{2}\int dx V(x,t)\frac{\partial}{\partial t}\langle \Omega_S(t)|\hat{\varphi}_S(x)^2|\Omega_S(t)\rangle \qquad (6.9)$$

Let $V(x,t)$ be given by,

$$V(x,t) = f(t)\lambda V(x) \qquad (6.10)$$

where $V(x)$ is the inverse step potential given by (1.3) and,

$$f(t) = \begin{cases} 1 \text{ for } t < 0 \\ 1 - \alpha t \text{ for } 1/\alpha > t > 0 \\ 0 \text{ for } t > 1/\alpha \end{cases} \qquad (6.11)$$

For $t < 0$ the scalar potential is static. The kinetic energy density for this case was determined in Section 2. We are interested in the change in the total kinetic energy during the time period from zero to $t = 1/\alpha$. This is given by,

$$\Delta E_K \left(0 \to (1/\alpha)\right) = \int_0^{(1/\alpha)} dt \frac{d}{dt} E_K(t) \tag{6.12}$$

For $t > 1/\alpha$, $dE_K(t)/dt = 0$ since $f(t) = 0$ for $t > 1/\alpha$.

At this point we will assume that the scalar potential is sufficiently small that we call solve this problem using standard perturbation theory as discussed in Chapt. 4 of Ref. [14]. In this case we re-write the Hamiltonian as,

$$\hat{H}_S = \hat{H}_{S0} + \hat{H}_{Sp} \tag{6.13}$$

where,

$$\hat{H}_{S0} = \hat{H}_{SKE} + \frac{1}{2} \int \lambda V(x) \hat{\varphi}_S(x)^2 \, dx \tag{6.14}$$

and,

$$\hat{H}_{Sp} = \frac{1}{2} \int \left(f(t) - 1\right) \lambda V(x) \hat{\varphi}_S(x)^2 \, dx \tag{6.15}$$

$\hat{H}_{S0}$ is the "unperturbed" part of the Hamiltonian and is calculated for a scalar field in the presence of the static scalar potential given by (1.3) and $\hat{H}_{Sp}$ is the part of the Hamiltonian that is due to the perturbation caused by the change in the scalar potential for $t > 0$. The Schrödinger picture field operators are given by,

$$\hat{\varphi}_S(x) = \hat{\varphi}_\lambda(x, 0) \text{ and } \pi_S(x) = \frac{\partial}{\partial t} \hat{\varphi}_\lambda(x, 0) \tag{6.16}$$

were $\hat{\varphi}_\lambda(x, t)$ is given by (2.4).

At this point we will follow standard perturbation theory as presented in [14] and switch to the Interaction picture where the interaction field operator is $\hat{\varphi}_\lambda(x, t)$ and the Interaction state vector is $|\Omega_I(t)\rangle$ where [14],

$$|\Omega_I(t)\rangle = e^{i\hat{H}_{S0}t} |\Omega_S(t)\rangle \text{ and } \hat{\varphi}_\lambda(x, t) = e^{i\hat{H}_{S0}t} \hat{\varphi}_\lambda(x) e^{-i\hat{H}_{S0}t} \tag{6.17}$$

In terms of Interaction picture quantities Eq. (6.9) becomes,

$$\frac{d}{dt} E_K(t) = -\frac{1}{2} \int V(x, t) \frac{\partial}{\partial t} \langle \Omega_I(t) | \hat{\varphi}_\lambda(x, t)^2 | \Omega_I(t) \rangle \tag{6.18}$$

The initial state vector, at $t = 0$, is $|\Omega_I(0)\rangle = |0_\lambda\rangle$. For $t > 0$ we have, to the lowest order in perturbation theory [14],

$$|\Omega_I(t)\rangle = \left(1 - i\hat{O}_1(t)\right)|0_\lambda\rangle; \quad \langle\Omega_I(t)| = \langle 0_\lambda|\left(1 + i\hat{O}_1(t)\right) \tag{6.19}$$

where,

$$\hat{O}_1(t) = \int_0^t \hat{H}_p(t) dt \tag{6.20}$$

$\hat{H}_p(t)$ is the perturbed part of the Hamiltonian in the Interaction picture and is given by,

$$\hat{H}_p(t) = \frac{1}{2}\int (f(t) - 1)\lambda V(x) \hat{\varphi}_\lambda(x,t)^2 dx \tag{6.21}$$

Use (6.20) to obtain,

$$\langle\Omega_I(t)|\varphi_\lambda(x,t)^2|\Omega_I(t)\rangle = i\langle 0_\lambda|\left[\hat{O}_1(t), \varphi_\lambda(x,t)^2\right]|0_\lambda\rangle \tag{6.22}$$

where the higher order terms have been dropped from the right hand side of this equation. Use this result in (6.18) to yield,

$$\frac{dE_K(t)}{dt} = -if(t)\int \lambda V(x) \frac{\partial}{\partial t}\langle 0_\lambda|\left[\hat{O}_1(t), \hat{\varphi}_\lambda(x,t)^2\right]|0_\lambda\rangle dx \tag{6.23}$$

This equation is valid to the lowest order in perturbation theory. It is assumed that the scalar potential is sufficiently small that the higher order terms can be ignored. Next we obtain,

$$\frac{\partial}{\partial t}\langle 0_\lambda|\left[\hat{O}_1(t), \hat{\varphi}_\lambda(x,t)^2\right]|0_\lambda\rangle = \langle 0_\lambda|\left[\hat{H}_p(t), \hat{\varphi}_\lambda(x,t)^2\right] + \left[\hat{O}_1(t), \frac{\partial\hat{\varphi}_\lambda(x,t)^2}{\partial t}\right]|0_\lambda\rangle \tag{6.24}$$

When this result is used in (6.23) the term involving $\left[\hat{H}_p(t), \hat{\varphi}_\lambda(x,t)^2\right]$ will end up being zero so that (6.23) becomes,

$$\frac{dE_K(t)}{dt} = -if(t)\int \lambda V(x) \langle 0_\lambda|\left[\hat{O}_1(t), \frac{\partial\hat{\varphi}_\lambda(x,t)^2}{\partial t}\right]|0_\lambda\rangle dx \tag{6.25}$$

At this point it won't affect the final result and will simplify the discussion if we write $\hat{\varphi}_\lambda(x,t)^2$ in normal order. Doing this and using (2.4) and (2.5) we obtain,

$$:\varphi_\lambda(x,t)^2:|0_\lambda\rangle = \sum_{\chi_1,\chi_2} \int_0^\infty dk \int_0^\infty dq \frac{\psi_k^{\chi_1}(x)\psi_q^{\chi_2}(x)}{2\pi\sqrt{kq}} e^{i(k+q)t} \hat{a}_k^{\chi_1\dagger}\hat{a}_q^{\chi_2\dagger}|0_\lambda\rangle \qquad (6.26)$$

From this it follows that,

$$\int \lambda V(x):\varphi_\lambda(x,t)^2:dx|0_\lambda\rangle = \sum_{\chi_1,\chi_2} \int_0^\infty dk \int_0^\infty dq \frac{C_{k,q}^{\chi_1\chi_2}}{2\pi\sqrt{kq}} e^{i(k+q)t} \hat{a}_k^{\chi_1\dagger}\hat{a}_q^{\chi_2\dagger}|0_\lambda\rangle \qquad (6.27)$$

and,

$$\langle 0_\lambda| \int \lambda V(x):\varphi_\lambda(x,t)^2:dx = \langle 0_\lambda| \sum_{\chi_1,\chi_2} \int_0^\infty dk \int_0^\infty dq \frac{C_{k,q}^{\chi_1\chi_2}}{2\pi\sqrt{kq}} e^{-i(k+q)t} \hat{a}_k^{\chi_1}\hat{a}_q^{\chi_2} \qquad (6.28)$$

where,

$$C_{k,q}^{\chi_1\chi_2} = \int \lambda V(x)\psi_k^{\chi_1}(x)\psi_q^{\chi_2}(x) dx \qquad (6.29)$$

From the above relationships we obtain,

$$\frac{\partial}{\partial t}\int \lambda V(x):\varphi_\lambda(x,t)^2:dx|0_\lambda\rangle = i\sum_{\chi_1,\chi_2} \int_0^\infty dk \int_0^\infty dq \frac{(k+q)C_{k,q}^{\chi_1\chi_2}}{2\pi\sqrt{kq}} e^{i(k+q)t} \hat{a}_k^{\chi_1\dagger}\hat{a}_q^{\chi_2\dagger}|0_\lambda\rangle \qquad (6.30)$$

and,

$$\langle 0_\lambda|\hat{O}_1(t) = \langle 0_\lambda| \frac{1}{2}\int_0^t dt'(f(t')-1) \sum_{\chi_1,\chi_2} \int_0^\infty dk \int_0^\infty dq \frac{C_{k,q}^{\chi_1\chi_2}}{2\pi\sqrt{kq}} e^{-i(k+q)t'} \hat{a}_k^{\chi_1}\hat{a}_q^{\chi_2} \qquad (6.31)$$

Use these results to obtain,

$$\langle 0_\lambda|\hat{O}_1(t)\frac{\partial}{\partial t}\int \lambda V(x):\varphi_\lambda(x,t)^2:dx|0_\lambda\rangle = i\sum_{\chi_1,\chi_2} \int_0^\infty dk \int_0^\infty dq \frac{(k+q)\left(C_{kq}^{\chi_1\chi_2}\right)^2}{4\pi^2 kq} B_{kq}(t) e^{i(k+q)t}$$

$$(6.32)$$

where,

$$B_{kq}(t) = \int_0^t dt'(f(t')-1)e^{-i(k+q)t'} \qquad (6.33)$$

Use these results in (6.25) to obtain,

$$\frac{dE_K(t)}{dt} = f(t) \sum_{\chi_1,\chi_2} \int_0^\infty dk \int_0^\infty dq \frac{(k+q)\left(C_{kq}^{\chi_1\chi_2}\right)^2}{4\pi^2 kq}\left(B_{kq}(t)e^{i(k+q)t} + B_{kq}(t)^* e^{-i(k+q)t}\right) \quad (6.34)$$

Use (6.11) in (6.33) to obtain for the time period $(1/\alpha) > t > 0$,

$$B_{kq}(t) = -\alpha \left\{ \frac{+i(k+q)t+1}{(k+q)^2} e^{-i(k+q)t} - \frac{1}{(k+q)^2} \right\} \tag{6.35}$$

Therefore,

$$\left( B_{kq}(t) e^{i(k+q)t} + B_{kq}(t)^* e^{-i(k+q)t} \right) = \frac{-2\alpha}{(k+q)^2} \left[ 1 - \cos\left[(k+q)t\right] \right] \tag{6.36}$$

Use this in (6.34) to obtain,

$$\frac{dE_K(t)}{dt} = -2\alpha(1-\alpha t) D(t) \tag{6.37}$$

where,

$$D(t) = \sum_{\chi_1,\chi_2} \int_0^\infty dk \int_0^\infty dq \, \frac{\left(C_{kq}^{\chi_1 \chi_2}\right)^2 \left(1 - \cos\left[(k+q)t\right]\right)}{4\pi^2 (k+q) kq} \tag{6.38}$$

It is evident that $D(t)$ is non-negative for all $t$. Therefore $dE_K(t)/dt \leq 0$ for all $t$ in the range $(1/\alpha) > t > 0$. We can use this result in (6.12) to show that $\Delta E_K(0 \to (1/\alpha))$ is negative.

How does this impact on the kinetic energy density. During the time interval $(1/\alpha) > t > 0$ the scalar potential is removed. This scalar potential was confined to the region $|a| > x$. The removal of this potential creates a disturbance which moves out at the speed of light from the region $|a| > x$. At $t = 1/\alpha$ the kinetic energy density in the region $|x| > a + (1/\alpha)$ is unaffected and, according to the results of Section 2, equals zero. The total kinetic energy in the region $|x| < a + (1/\alpha)$ has decreased by the amount $\left| \Delta E_K (0 \to (1/\alpha)) \right|$ from $t = 0$ to $t = 1/\alpha$. From the results of Section 2 the total kinetic energy within this region at $t = 0$ was $E_{KE}$ which is negative.

Therefore the situation at $t = 1/\alpha$ is this – the kinetic energy density for $|x| > a + (1/\alpha)$ is zero. The total kinetic energy in the region $|x| < a + (1/\alpha)$ is $E_{KE} + \Delta E_K (0 \to (1/\alpha))$ where both terms are negative. Proceeding as in Section 4 it is evident that the spatial quantum inequality is violated at $t = 1/\alpha$.

## 7. The quantum interest conjecture.

At this point we have specified the field operator and kinetic energy density at $t = 0^+$. What happens for $t > 0^+$. The field operator evolves in time according to Eq. (3.1). A similar problem was considered in Ref. [8]. Using the same analysis as in [8] we can show that,

$$T_{00R}^{(+)}(x,t) = \frac{1}{2}\left(T_{00R}(x+t) + T_{00R}(x-t)\right) \tag{7.1}$$

where $T_{00R}(x)$ is the kinetic energy density calculated in Section 2.

This result means that for $t > 0$ we have two pulses of negative energy moving in opposite directions at the speed of light. The pulses are isolated. That is they are not surrounded or associated with a positive energy pulse. This is a violation of the quantum interest conjecture.

The quantum interest conjecture states that a pulse of negative energy must be preceded or followed by an even larger pulse of positive energy [13]. However, based on the results from Ref. [8], we have a pulse of negative energy that is not associated with a pulse of positive energy. Therefore the quantum interest conjecture fails.

## 8. Conclusion.

We have examined a scalar field with zero mass in 1-1 dimensional space-time subject to the scalar potential $V(x)$ given in (1.3). We have shown that the kinetic energy density is negative inside the region $x < |a|$ and zero outside of this region. When the scalar potential is instantaneously removed at $t = 0$ we can determine the energy density for $t = 0^+$ by using the fact that the kinetic energy density is continuous at $t = 0$ despite the instantaneous change in the scalar potential. From this we are able to show that the spatial quantum inequality is violated.

## Appendix.

Use (2.32) and (2.33) in (2.31) and let $\lambda = 1$ to obtain,

$$E_{KE} = -\frac{1}{4\pi} \int_0^\infty d\kappa \left[ \begin{array}{l} \left( \dfrac{2\kappa^2}{\kappa'} + \dfrac{V_0}{\kappa'} \right) \left( \dfrac{2aN_k + V_0 \sinh(2a\kappa')/\kappa'}{D_\kappa} \right) \\ -4\kappa a - \dfrac{2V_0 \sinh(2a\kappa')}{D_\kappa} \end{array} \right] \quad (9.1)$$

Rearrange terms to obtain,

$$E_{KE} = -\frac{1}{4\pi} \int_0^\infty \frac{d\kappa}{D_\kappa} \left[ 2aN_k \left( \frac{2\kappa^2 + V_0}{\kappa'} \right) - 4\kappa a D_\kappa + V_0 \left( \frac{2\kappa^2 + V_0}{\kappa'^2} - 2 \right) \sinh(2a\kappa') \right] \quad (9.2)$$

Use $\kappa' = \sqrt{\kappa^2 + \lambda V_0}$ in the above to obtain,

$$E_{KE} = -\frac{1}{4\pi} \int_0^\infty \frac{d\kappa}{D_\kappa} \left[ 2aN_k \left( \frac{2\kappa'^2 - V_0}{\kappa'} \right) - 4\kappa a D_\kappa - \frac{V_0^2}{\kappa'^2} \sinh(2a\kappa') \right] \quad (9.3)$$

Rearrange terms to obtain,

$$E_{KE} = -\frac{1}{4\pi} \int_0^\infty \frac{d\kappa}{D_\kappa} \left[ 4a(\kappa' N_k - \kappa D_\kappa) - \frac{2aN_k V_0}{\kappa'} - \frac{V_0^2}{\kappa'^2} \sinh(2a\kappa') \right] \quad (9.4)$$

Use (2.28) and (2.29) in the above to obtain,

$$E_{KE} = -\frac{1}{4\pi} \int_0^\infty \frac{d\kappa}{D_\kappa} \left\{ \begin{array}{l} \cosh(2a\kappa') \left[ 4a\kappa'(\kappa'^2 - \kappa^2) - \dfrac{2aV_0}{\kappa'}(\kappa'^2 + \kappa^2) \right] \\ + \sinh(2a\kappa') \left[ 4a\kappa(\kappa'^2 - \kappa^2) - 4a\kappa V_0 - \dfrac{V_0^2}{\kappa'^2} \right] \end{array} \right\} \quad (9.5)$$

Rearrange terms and use $\kappa' = \sqrt{\kappa^2 + \lambda V_0}$ with $\lambda = 1$ to obtain Eq. (2.34) in the text.

**References.**